\documentclass{emulateapj} \usepackage{psfig} \usepackage{apjfonts}

\newcommand\beq{\begin{equation}}
\newcommand\eeq{\end{equation}}

\begin{document}
\title{Flare-less long Gamma-ray Bursts and the properties of their 
massive star progenitors}

\author{Rosalba Perna\altaffilmark{1} and Andrew MacFadyen\altaffilmark{2}} 
\altaffiltext{1}{JILA and Department of Astrophysical and Planetary Sciences, 
University of Colorado, Boulder, CO, 80309}
\altaffiltext{2}{Center for Cosmology and Particle Physics, Department of Physics, 
New York University, 4 Washington Place, New York, NY 10003}

\begin{abstract}

  While there is mounting evidence that long Gamma-Ray Bursts (GRBs)
  are associated with the collapse of massive stars, the detailed
  structure of their pre-supernova stage is still
  debatable. Particularly uncertain is the degree of mixing among
  shells of different composition, and hence the role of magnetic
  torques and convection in transporting angular momentum. Here we
  show that early-time afterglow observations with the {\em Swift}
  satellite place constraints on the allowed GRB pre-supernova
  models. In particular, they argue against pre-supernova models in
  which different elemental shells are unmixed. These types of models
  would produce energy injections into the GRB engine on timescales
  between several hundreds of seconds to a few hours. Flaring activity
  has {\em not} been observed in a large fraction of well-monitored
  long GRBs. Therefore, if the progenitors of long GRBs have common
  properties, then the lack of flares indicates that the massive stars
  which produce GRBs are mostly well mixed, as expected in
  low-metallicity, rapidly rotating massive stars.
 
\end{abstract}


\section{Introduction}

Theoretical studies (e.g. MacFadyen \& Woosley 1999) have suggested
that long GRBs are produced by the collapse of rapidly-rotating,
low-metallicity massive stars. In the last few years, the association
between a GRB and the death of a massive star has found strong support
through observations of supernovae associated with GRBs (Stanek et
al. 2003; Hjorth et al. 2003).  

Despite this progress, many details of the evolution of the massive
star (and hence on its inner structure at the moment of death) are
still rather uncertain. Attempts to constrain the characteristics of the
GRB progenitors have been made by Campana et al. (2008). They performed
a detailed study of the X-ray spectrum of GRB~060218, discovering a lower
than normal O/N ratio in the surrounding of the burst progenitor. They
concluded that only a progenitor star characterized by a fast stellar
rotation and sub-solar initial metallicity could produce such a metal
enrichment. 

What remains very uncertain however is the interior structure of the
pre-supernova star, and in particular the extent to which the
different burning layers are mixed. This has important implications
for the physics of angular momentum transport in the stellar interior,
and for the relevance of magnetic fields in the stellar evolution.  No
direct observational constraints have been obtained so far on such
quantities.

In this {\em Letter} we identify an observational signature in the GRB
phenomenology which bears direct consequences for the structure of the
pre-supernova star.  The launch of the {\em Swift} satellite in
November 2004 has opened a new window of early-time observations of
Gamma-Ray Bursts (GRBs) and their afterglows.  Early-time observations
with the {\em Swift} X-ray Telescope (XRT) telescope have revealed a
new and unexpected phenomenology. Nearly half of the bursts displayed
erratic X-ray flares peaking on timescales between several hundreds of
seconds to several hours (Burrows et al. 2005; Nousek et al. 2006;
O'Brien et al. 2006; Falcone et al. 2006, 2007; Chincarini et
al. 2007).  The observed phenomenology has been found to be common to
both long and short GRBs, and a number of studies have been aimed at
establishing the origin of these X-ray rebrightenings. While in some
cases the data are consistent with a refreshed (external) shock
(Guetta et al. 2007), the greater majority of the flares do require an
extended activity of the inner engine, due to their fast temporal rise
and decay (Zhang et al. 2006; Lazzati \& Perna 2007; Chincarini et
al. 2007; Falcone et al. 2007) and as indicated by internal shock
modeling (Maxham \& Zhang 2009).

Several authors have explored ideas for producing a continued activity
of the GRB inner engine after the time frame of the classical prompt
emission. King et al. (2005) suggested that the X-ray flares could be
produced by means of the fragmentation of the collapsing stellar core
in a modified collapsar scenario. Dai et al. (2006) proposed that the
flares can be created by magnetic reconnection events driven by the
breakout of magnetic fields from the surface of differentially
rotating millisecond pulsars produced from a binary merger. An engine
that can be stopped and restarted following repeated episodes of
magnetic flux accumulation and release was hypothesized by Proga \&
Zhang (2006). The observational properties of the flares, namely
a correlation between their duration and their arrival times and
an anticorrelation between their duration and their peak luminosity,
together with their common existence in both long and short GRBs, 
prompted Perna et al. (2006) to notice that the origin of the flares 
is consistent with viscous evolution of several rings of material, which
they suggested were produced by fragmentation or large-amplitude
variability within the hyperaccreting inner disk.

Whichever the underlying reason for the flares, here we focus our
attention on two important facts: {\em (a)} flares are observed in
both long and short GRBs, and with similar properties; {\em (b)} more
than half of the well-monitored bursts do {\em not} display any
flaring activity\footnote{Out of the 110 {\em Swift} bursts that
  Falcone et al (2007) studied, they found that only $33\%$ of them
  had evidence for flares at more than the $3\sigma$ level.}  (this is
true independently for both the long and the short class).  We
interpret these facts as a hint that the flaring activity is unlikely
to be directly related to the progenitor structure\footnote{
In the context of this work, we indicate by 'progenitor' the 
astrophysical entity {\em prior} to the formation of a disk.}. In fact, if long
and short GRBs have different types of progenitors, as commonly
believed, then it would be hard to explain point {\em
  (a)}. Furthermore, if the progenitors of each class have similar
properties, then is it not straightforward to account for {\em (b)}.
Flares are most likely related to a common element of long and short
GRBs, whose properties can display a large degree of variation from
case to case.

With the above considerations in mind, we will argue here that the
lack of flares in a substantial fraction of long GRBs places
constraints on the interior structure of the pre-supernova
star progenitor of long GRBs. 

The paper is organized as follows: \S 2 describes the main features of
the collapsar model which are critical for our discussion. These are
related in \S 3 to the GRB phenomenology. Finally, we summarize
in \S 4.

\section{The angular momentum structure of the pre-supernova star 
in different models}

The collapsar model (Woosley 1993; MacFadyen \& Woosley 1999;
MacFadyen, Woosley \& Heger 2001, MWH01 in the following) for long
GRBs has received strong support by the observation of supernovae
associated with this class of GRBs (Stanek et al. 2003; Hjorth et
al. 2003). The iron core of a rapidly-rotating star of mass greater
than $\approx 25M_\odot$ collapses to form a black hole at its
center. The outer layers fall toward the newly formed black hole and
are halted at the Kepler radius.  The location of the circularization
radius of the infalling material contains direct information on the
angular momentum structure of the pre-supernova star, and hence on the
mechanisms by which angular momentum is transferred in the interior of
the evolving star.

Earlier studies of massive stellar evolution (with the exception of a
few, such as Spruit \& Phinney 1998 and Maeder \& Meynet 2004), did
not include a possibly important effect, that is the torque exerted in
differentially rotating regions by the magnetic field that threads
them.  {With all the uncertainties regarding the correct treatment
of $B$-field generation and evolution (e.g. Rudiger \& Hollerbach
2004),} a first attempt to include angular momentum transport by
magnetic torques in simulations of pre-supernova stars was made by
Heger et al. (2005). They noted, however, that their results are quite
sensitive to several uncertain model parameters, such as the
efficiency of the dynamo generating the magnetic field, the effect of
composition gradients, and the initial angular momentum of the star.
{It should also be pointed out that the standard diffusion
treatment of angular momentum transport in radiative layers adopted by
these authors (as well as others), fails to reproduce some basic
features, such as the solar rotation profile and the Li dip in F
stars.  It is also at odds with helioseismic inversions (Thompson et
al. 2003) and 3D numerical simulations (Miesch et
al. 2006)\footnote{A substantial improvement between theory and
observations has been made by the inclusion of the transport of
angular momentum by internal waves in stellar interiors (Zhan et
al. 1997; Talon et al. 2002; Charbonnel \& Talon 2005). }}.
Furthermore, and of special importance here, the inclusion of angular
momentum transport by magnetic torques during the precollapse
evolution is known to create problems for those stars that would
become GRBs. In fact, while GRBs are believed to be produced by stars
endowed with a particularly large rotation, the angular momentum
transport saps the core of the necessary rotation.  This problem was
addressed by Woosley \& Heger (2006; see also Yoon \& Langer
2005). They studied the evolution of very rapidly rotating stars, and
found that, for the highest possible rotation rate, a new evolutionary
channel appears, in which single stars fully mix while still on the
main sequence, and never become red giants. Under specific conditions,
these stars (which comprise about 1\% of the massive star population),
could retain enough angular momentum to produce GRBs. In this
scenario, the angular momentum in the interior of the pre-supernova
star has a continuous distribution with radius, unlike the case in
which the shells of the various burning elements are unmixed. {A
rather strong compositional mixing is also obtained in 2D and 3D
numerical simulations of pre-SN massive star interiors that include
long-distance angular momentum redistribution and convection-induced
mixing by the excitation of internal waves (Meakin \& Arnett 2006;
2007; see also Mocak et al. 2009 for similar findings in the context
of He flash driven convection).}

Given all of the uncertainties still present in the current
understanding of the progenitor stars of GRBs, being able to validate
the various theoretical models by means of direct observational tests
is very valuable.

In the following, we will consider the observational consequences
for the GRB phenomenology of different types of GRB progenitor
stars.

\begin{figure*}[t]
\parbox{0.95\textwidth}{
\psfig{file=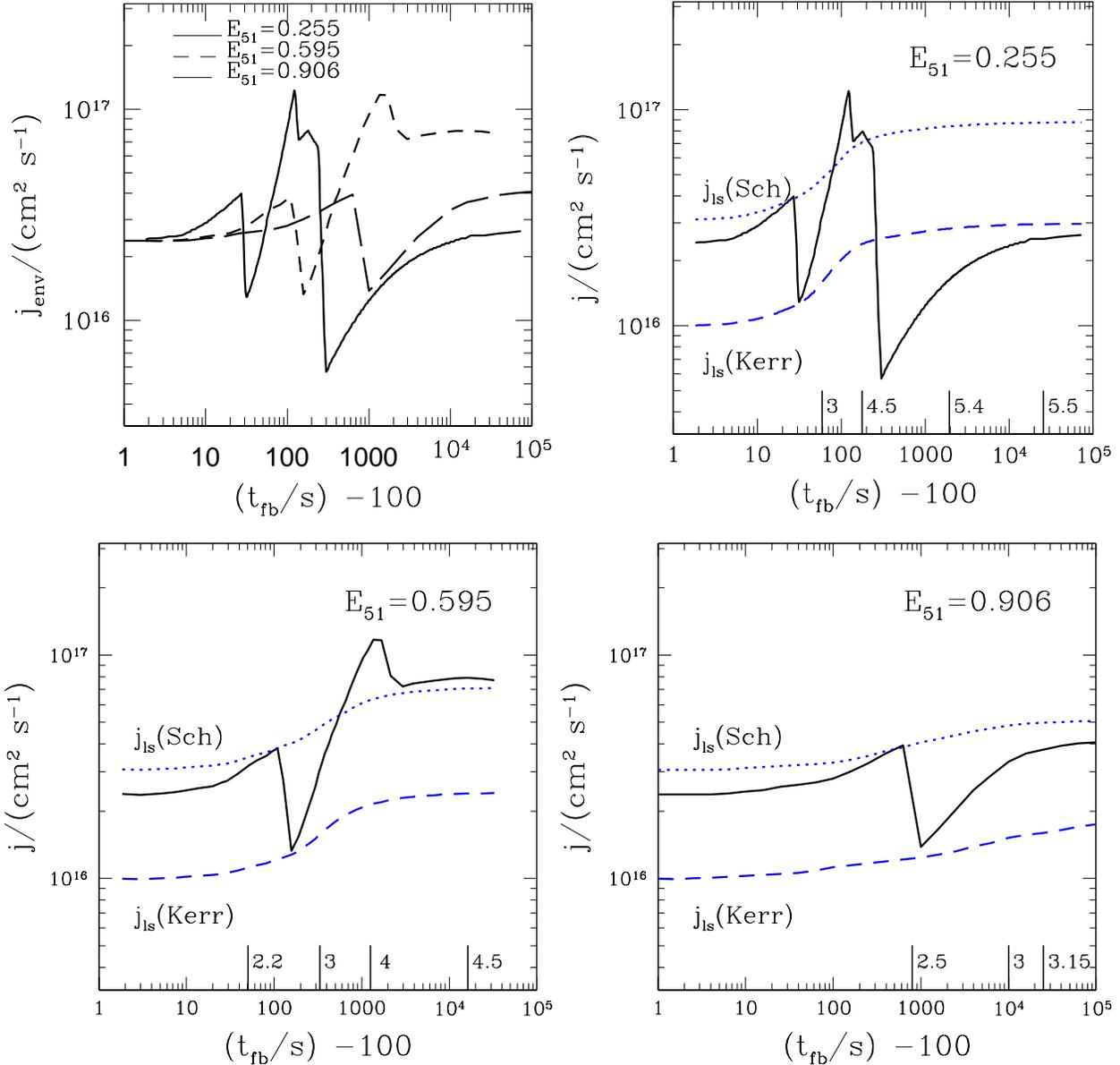,angle=0,width=0.95\textwidth}}
%\plotone{fig.ps}

\caption{The specific  angular momentum of  the burning shells  in the
  envelope  of the pre-supernova  star studied  in MWH01  versus their
  fallback  time  $t_{\rm fb}$  after  the  explosion.  The  structure
  results from the evolution of  a star of mass $M=10^{25}M_\odot$ and
  an equatorial  rotational velocity  on the main  sequence of  200 km
  s$^{-1}$.  This  model assumes  no mixing between  different burning
  layers. The top  left panel compares the envelope  fallback time for
  three different explosion energies.  The other three panels compare,
  for  each  value  of  the  explosion energy,  the  specific  angular
  momentum  of  the shells  to  the  specific  angular momentum  of  a
  particle  at  the  last  stable  orbit,  in  the  two  limits  of  a
  Schwarzschild and  of a  Kerr black hole.   The numbers next  to the
  vertical ticks indicate the amount of mass accreted up to that time,
  in units of  solar mass. When $t_{fb}=100$~s, a black  hole of mass of
  $2M_\odot$ has already formed,  partly from direct core collapse and
  partly  from subsequent  accretion.   For a  rapidly rotating  black
  hole, as expected from the collapse of the core of the pre-supernova
  star  under  consideration,   distinct  episodes  of  accretion  are
  expected, with  the last  extending to timescales  of up  to several
  hours. }
\vspace{0.1in}
\label{fig:fig1}
\end{figure*}

\section{Implications for GRB phenomenology of the pre-supernova
interior structure}

If the star is fully mixed, as suggested by Woosley \& Heger (2006),
then the angular momentum is a continuous function of the radius $r$,
which can be approximated as (Kumar, Narayan \& Johnson 2008a) \beq
j(r)\approx 3.8\times 10^{18}\, M_{\rm BH}^{1/2}\,r_{10}^{1/2}
f_\Omega(r)\;\; {\rm cm}^2\, {\rm s}^{-1}\,,
\label{eq:angm}
\eeq
where $r_{10}=r/(10^{10} \;{\rm cm})$, $M_{\rm BH}$ is the black hole
mass in units of solar masses, and $f_\Omega$ is the ratio between
the angular velocity of the gas and the local Keplerian angular
velocity. This model, as demonstrated by Kumar et al. (2008a; 2008b),
can produce plateaus in the afterglow light curve, but the
overall trend of the time evolution of the afterglow is generally smooth.

In the following, we will examine the consequences for the GRB
phenomenology of the unmixed GRB progenitor models, and we will
consider, as a representative example, the pre-supernova star model
used by MWH01, with mass $M=25 M_\odot$, and an equatorial rotational
velocity on the main sequence of 200 km s$^{-1}$. While the specific
details of our results are expected to vary depending on the precise
location and angular momentum of the shells, our general arguments are
expected to hold.

In these types of pre-supernova stars, the burning layers of different
elements are unmixed, and the specific angular momentum of the various
shells is different, increasing with radius.  Ejected material that
does not possess enough energy to unbind falls back, eventually
settling at the circularization radius, depending on the value of its
specific angular momentum $j$.  While the iron core collapses to form a
black hole, the outer burning shells receive a radial momentum which
results in some ejecta becoming unbound.  The bound material falls
back, eventually settling into a ring at the Keplerian radius. From
that point on, the evolution is driven by viscosity and by accretion
of low angular momentum material from the polar regions.

The innermost material, close to the last stable orbit, accretes rapidly
at high rates, giving rise to the prompt emission, while
the  outer shells start accreting at later times. 
There are two times which combine to produce the timescales 
over which the rebrightenings due to accretion from the outer shells take place.
The first is the time that the ejecta take to fall back to the circularization
radius after being propelled outwards. A hard limit to this is given by the
free fall time 
\beq
t_{ff}\sim \frac{1}{\sqrt{G\rho_{ave}}},
\label{eq:tff}
\eeq where $\rho_{ave}$ is the average density of the material inside
the shell under consideration.  Even for the outermost shell, the
dynamical time is on the order of seconds.  A more realistic estimate
of the fallback time must include the effects of the kinetic energy
(and hence the radial velocity) that is imparted to the envelope by
the explosion. The mass fallback rate for a range of explosion
energies between $0.255\times 10^{51}$ erg and $1.682\times 10^{51}$
erg was computed for the model star under consideration by MWH01.  In
the equatorial region, which is of interest for our study, the
effective energy is expected to be in the lower range.

The specific angular momentum of the envelope, $j_{\rm env}$, as a
function of the fallback time is displayed in Fig.1 for the effective
kinetic explosion energies $E=0.255\times 10^{51}$ erg, $E=0.595\times
10^{51}$ erg, and $E=0.906\times 10^{51}$ erg (see top left panel). In
order for a disk to form, $j_{\rm env}$ must be larger than the
specific angular momentum $j_{\rm ls}$ of a particle orbiting at the
last stable orbit.  The specific angular momentum of a particle on a
corotating orbit of a black hole of mass $M$ and angular momentum
$J=aM$ is given by \beq j = \frac{\sqrt{GMR} \left[R^2 -
    2(a/c)\sqrt{GMR/c^2} +(a/c)^2\right]} {R\left[ R^2-3GMR/c^2 +
    2(a/c)\sqrt{GMR/c^2}\right]^{1/2}}\;.
\label{eq:jGR}
\eeq The right top panel and the bottom panels of Fig.1 respectively
show, for each of the three considered explosion energies, $j_{\rm
  env}$ as compared to the minimum specific angular momentum $j_{\rm
  ls}$ needed to form a disk at the last stable orbit, for the
limiting cases of a Schwarzschild ($a=0$) and of a Kerr ($a=GM/c$)
black hole.  A minimum value of $j_{\rm ls}$ (and hence of the Kerr
parameter), for each explosion energy, can be derived by the condition
that, at early times, $j_{\rm env}>j_{\rm ls}$, or else there would
not be a disk to power the prompt GRB phase. From Fig.1 it can be seen
that, if the innermost material has enough angular momentum to
circularize in a disk, this will also be the case for an outer shell
of the envelope.  The precise location of the circularization radius
will depend on the specific angular momentum of the material, as well
as on that of the black hole, according to Eq.(\ref{eq:jGR}).  For the
outermost shell, we find the circularization radius to be on the order
of about $10\,R_s$.  Once the bound material has circularized in a
ring, the following evolution occurs on the viscous timescale \beq
t_{\rm visc}(R_{\rm circ})=\frac{R_{\rm circ}^2}{H^2 \alpha \Omega_K}
\sim 5\times 10^{-4} \alpha^{-1}_{-1}m_3
r^{3/2}\left(\frac{R}{H}\right)^{2}\;{\rm s} \;,
\label{eq:t0}
\eeq
where $\Omega_K$ is the Keplerian
velocity of the gas in the disk, $H$ the disk scale-height,  
$\alpha$ the viscosity parameter (Shakura \& Sunyaev 1973), and 
$r$ the radius in units of the Schwarzschild's radius. 
The specific functional form of $t_{\rm visc}$ with radius depends on the
properties of the disk. 
 
If advection dominates, which is likely to be the case for very large
accretion rates, then the disk scale height is $H \sim R$, and the
viscous timescale at $r\sim 25$ where the outermost shell circularizes
is $\sim 0.04$ sec.  Once a new, outer ring of material catches up
with the inner accreting ring, a secondary episode of accretion is
likely to occur.  A discussion of this enhanced, late energy output
was presented by Lee, Ramirez-Ruiz \& Lopez-Camara (2009) for the case
in which the disk is produced by the debris of the tidal disruption of
a compact object in a binary merger, and the late accretion episode
derives from material further out in the tidal tail.  They found that
the injection of this material can re-energize the main accretion
disk, as long as the fallback mass dominates over the remnant disk
mass at the time at which it is re-injected at its circularization
radius. The following evolution then proceeds on the viscous time of
the resulting ring. Their simulations also showed that the secondary
episode of accretion produces a total energy output and neutrino
luminosity which are comparable to those observed in flares.

{In the pre-supernova model considered here we would expect, at a
qualitative level, a similar phenomenology. Given the specific angular
momenta of the two shells, and the corresponding BH mass at the time they accrete,
from inversion of Eq.(\ref{eq:jGR}), we find that the ratio between
the circularization radii of the two shells varies between $R_{\rm
circ}^{\rm out}/R_{\rm circ}^{\rm in}\sim 8.6$ for $a=0.2\,GM/c$, and
$\sim 5.6$ for $a=GM/c$. This, together with the fact that $M_{\rm
out}^{\rm shell}\sim 4\, M_{\rm in}^{\rm shell}$, implies that, once
the material circularizes, the accretion rates of the two shells are
expected to be comparable.  Hence, since the fallback time of the
outer shell is $t_{\rm fb}^{\rm out}\ga 10\, t_{\rm fb}^{\rm in}\gg
t_{\rm visc}$, and the accretion rate of the inner shell drops as
$\dot{M}_{\rm in}\propto \left(t/t_{\rm visc}\right)^{-{5/3}}$ for
$t\ga t_{\rm fb}^{\rm in}$ (MWH01), the arrival of the outer shell on
the timescale $t_{\rm fb}^{\rm out}$ will produce a significant
enhancement in the tail of $\dot{M}_{\rm in}$. } Therefore, as long as the
accretion efficiency at late times is not suppressed with respect to
that at early times, we would expect that the late, massive outer
shells release a luminosity comparable to that in the
prompt phase.  The expected phenomenology in GRBs would hence be that
of powerful flares superimposed to the powerlaw decay of the afterglow, with
the specific number of flares and their arrival times depending on the
details of the pre-supernova structure (here we just considered one 
particular case).

Among all the long GRBs detected by {\em Swift} and well monitored
after the trigger, flares have been detected in about 30\% of the
cases. If, as discussed in \S1, we make the reasonable assumption that
the properties of the pre-supernova stars that produce long GRBs are
similar, then the lack of flares
in the largest majority of long GRBs  argues in favor of a
pre-supernova model which is fully mixed, as found in the simulations
by Woosley \& Heger (2006) for a low-metallicity, rapidly-rotating
star.
A well-mixed envelope might be a likely outcome also in the He-star/BH
merger scenario (Fryer \& Woosley 1998), in which the GRB is preceded
by a common envelope phase during which the BH enters first the H and then
the He envelope of the star. The stirring accompanying the merger process is
likely to wash out any shell structure within the star. 

\section{Summary}

While it has long been recognized that the collapse of a massive star
under particular circumstances can produce a long GRB, what are these
``particular circumstances'' is still a subject of exploration.  The
generally important role of a high angular momentum has been widely
discussed, but the detailed structure of the interior of the
pre-supernova star, connected to the extent to which angular momentum
transport is effective within its envelope, is still rather uncertain.

In this {\em Letter} we have shown that late energy injection, from
several hundreds of seconds to several hours, is a common feature of a
collapsar model in which the pre-supernova star has unmixed shells of
burning elements. The lack of flaring activity in more than half of
the well-monitored GRBs hence argues against such a model.  The
arguments presented here give support to a fully mixed pre-supernova,
as expected for a rapidly-rotating, low-metallicity massive star.

\acknowledgements We thank Abe Falcone, Bing Zhang and Weiqun Zhang
for discussions on various aspects of this work.

\end{document}